# Second-harmonic generation with a 440,000% W$^{-1}$ conversion efficiency in a lithium niobate microcavity without periodic poling

*Xiao Wu, Zhenzhong Hao, Li Zhang, Di Jia, Rui Ma, Fang Bo,\* Feng Gao, Guoquan Zhang, and Jingjun Xu*

X. Wu, Z. Hao, L. Zhang, D. Jia, R. Ma, F. Bo, * F. Gao, G. Zhang, * J. Xu*
MOE Key Laboratory of Weak-Light Nonlinear Photonics, TEDA Institute of Applied
Physics and School of Physics, Nankai University, Tianjin 300457, China
E-mail: bofang@nankai.edu.cn; zhanggq@nankai.edu.cn; jjxu@nankai.edu.cn

Keywords: second-harmonic generation, thin-film lithium niobate, phase matching, microcavity

Thin-film lithium niobate (TFLN) enables extremely high-efficiency second-order nonlinear optical effects due to large nonlinear coefficient $d_{33}$ and strong optical field localization. Here, we first designed and fabricated a pulley-waveguide-coupled microring resonator with an intrinsic quality factor above $9.4 \times 10^5$ on the reverse-polarized double-layer X-cut TFLN. In such a TFLN resonator without fine domain structures, second harmonic generation with an absolute (normalized) conversion efficiency of 30% (440,000% W$^{-1}$), comparable to that in periodically poled lithium niobate (PPLN) microring resonators, was realized with a sub-microwatt continuous pump. This work reduces the dependence of high-efficiency nonlinear frequency conversion on PPLN microcavities that are difficult to prepare.

## 1. Introduction

Ultra-efficient second harmonic (SH) generation based on second-order nonlinearity has essential applications in classical light sources[1-3] and quantum light sources.[4-6] Thin-film lithium niobate (TFLN) microring resonators with high-quality factors have been widely used in nonlinear photonic integration platforms due to their large nonlinear coefficient ($d_{33}$=25 pm V$^{-1}$), wide transparent window (0.4 ~ 5 μm) and tight light field localization.[7-9] The flexible microdomain processing of TFLN enables quasi-phase-matched periodically poled lithium niobate (PPLN) photonic waveguides and microcavities with extremely high nonlinear optical conversion efficiencies.[10-12] In 2021, Lei Wang et al. innovatively realized mode phase matching (MPM) from fundamental harmonic (FH) TE$_{00, FH}$ mode to second harmonic (SH)

TE$_{10, SH}$ mode in reverse-polarization (RP) dual-layer TFLN waveguides. The normalized SH conversion efficiency was estimated to be 5540% W$^{-1}$ cm$^{-2}$, comparable to the SH conversion efficiency of PPLN waveguides.[13-15] The main advantages of MPM on RP TFLN compared to PPLN are: (1) The effective nonlinearity coefficient is equal to $d_{33}$ instead of $2/\pi \cdot d_{33}$, although at the cost of smaller mode overlap. (2) Electrode preparation and subsequent rigorous periodic poling processes are not required.

Here, we designed and prepared pulley-waveguide-coupled racetrack microresonators on an RP double-layer X-cut TFLN. In these TFLN resonators without periodic domain structures, an ultra-high normalized conversion efficiency of 440,000% W$^{-1}$ was achieved under nearly critical pump coupling and dual FH and SH resonance. The normalized SH conversion efficiency exceeds that of all photonic microcavities based on MPM, only an order of magnitude lower than PPLN microcavities with complex preparation processes.[16] Furthermore, a 30% absolute conversion efficiency was observed with a 600-μW on-chip pump power.

## 2. Principles of Device Design and Fabrication

The racetrack resonators were fabricated on an RP double-layer TFLN on a 2 μm thick silicon dioxide film and 500 μm thick silicon substrate. The total thickness of the TFLN is 600 nm, and the upper layer has a thickness of 310 nm. The preparation method of TFLN microring resonators is similar to that in Reference.[17] It mainly includes electron beam lithography (EBL) to define the etching mask and inductively coupled plasma reactive ion etching (ICP-RIE) to transfer the mask pattern to the TFLN, where the complex electrical or optical poling processes to prepare the microdomain structure of PPLN devices is excluded. The scanning electron microscope (SEM) of a fabricated TFLN microring resonator is shown in **Figure 1**a. The length of the straight arm of the racetrack resonator is 300 μm. To reduce mode coupling between the TE$_{10, SH}$ mode and other high-order modes ensuring the adiabatic mode evolution in the bend section, the bending radius of the racetrack cavity was set to be 150 μm. We used a single pulley waveguide to couple and extract the pump and the signal. The width of the pulley waveguide is 450 nm, the coupling gap is 300 nm, and the coupling length is 52 μm. The zoomed SEM images of the coupling part and the cross-section are shown in Figure 1b and its inset. The smooth waveguide surface and side walls benefit from our high-quality fabrication process excluding periodic poling. On the cross-section, the dividing line of the RP double-layer film is faintly visible.

Figure 1c schematically illustrates the cross-section of the straight arm of the RP double-layer TFLN resonator, with crystal orientation as coordinates. The yellow arrows in the waveguide indicate the direction of spontaneous polarization of lithium niobate crystal. The etching depth of 420 nm allows $TE_{10, SH}$ a more minor mode loss. The phase matching condition is realized by adjusting the top width of the waveguide. We used the finite difference method to simulate the effective refractive indices $n_{eff}$ of $TE_{00, FH}$ mode around 1550 nm and $TE_{10, SH}$ mode near 775 nm in a straight waveguide as indicated in Figure 1d. The phase matching condition requires that the $n_{eff}$ of the two modes be equal.[18] Accordingly, the waveguide width is selected to be 1.21 µm, as shown at the upper cross point of the red and blue lines. In such a structure, the nonlinear coupling strength $g$ can reach 6.35 MHz ($g/2\pi$=1.01 MHz) (See Supporting Information). We prepared waveguides of different widths with steps of 30 nm to compensate for design and fabrication uncertainties, including the refractive index, film thickness, etching depth, and waveguide width, to achieve phase matching for SH generation in the TFLN resonator.

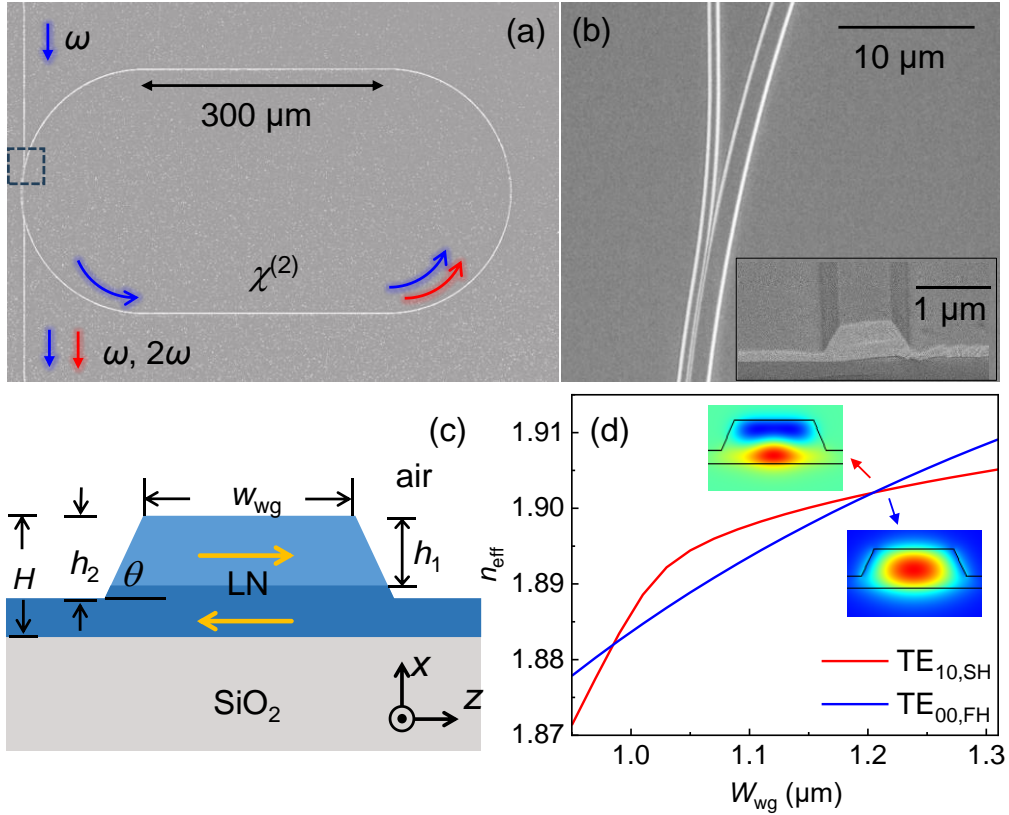

**Figure 1.** Design and characterization of double-layer TFLN racetrack resonators. a) SEM images of the overall structure of a typical TFLN racetrack ring. b) SEM image of the coupled region between a pulley waveguide and a resonator. The inset is the SEM image demonstrating the end face of the coupled waveguide. c) Schematic diagram of the cross-

section of the straight arm in a racetrack RP dual-layer resonator, where $H$=600 nm, $h_1$=310 nm, $h_2$=420 nm, $\theta$=65°. d) The effective refractive indexes of $TE_{00,\ FH}$ and $TE_{10,\ SH}$ modes vary with the waveguide width. The insets show the $E_z$ component of the corresponding modes.

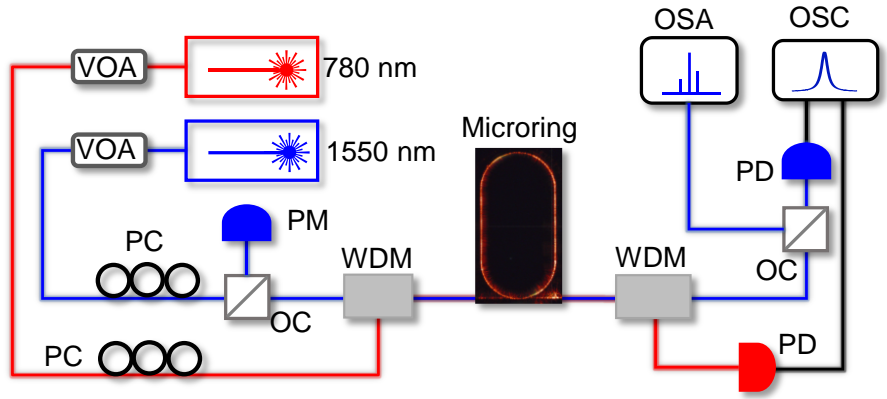

**Figure 2.** The experimental setup for SH detection. The racetrack resonator is a photograph that scatts the SH signal. VOA, variable optical attenuator; PC, polarization controller; WDM, wavelength division multiplexer; OC, optical coupler; PM, power meter; PD, photodetector; OSC, oscilloscope; OSA, optical spectrum analyzer.

## 3. Second Harmonic Generation

The setup shown in **Figure 2** was utilized to characterize the SH signal in the RP double-layer TFLN microcavity. Tunable lasers in the 1550 and 780 nm bands were combined by a wavelength division multiplexer and coupled to the chip through a lensed fiber. The power of the 1550 nm pump was controlled by a variable optical attenuation and monitored in real time by a coupler and an optical power meter. The polarization controller is used to excite the on-chip $TE_{00,\ FH}$ mode. The pump and the generated SH signal are coupled out of the chip through another lensed fiber, separated by a wavelength division multiplexer, detected by their respective photodetector, and then fed into the oscilloscope. The 780 nm laser is mainly used to estimate the coupling loss of the signal and, on this basis, to calibrate the signal power monitored by the photodetector and oscilloscope. Assuming the two end-coupling efficiencies are the same, the fiber-to-chip losses of the pump and SH estimated based on transmittance are 8.4 dB/facet and 8.0 dB/facet, respectively. The chip is placed on a stage with a thermoelectric controller and is temperature-adjusted to achieve double resonances of FH and SH to achieve maximum conversion efficiency. The microresonator in Figure 2 is a photograph that lights up due to the scattering of SH light.

The most efficient SH conversion was observed in a TFLN resonator with a 1.27 μm top width. Experimentally, pump transmission and SH signals were collected simultaneously while scanning the pump laser. Typical results are shown in **Figure 3**a, where the arrows in the transmission spectrum point to the TE$_{00, \text{FH}}$ mode, with a free spectral range (FSR) of 0.647 nm. The pump and signal modes that meet phase matching and energy conservation locate at 1518.4 nm and 759.2 nm, respectively. Optimizing polarization of pump light, the transmission spectrum of the pump and SH near the mode with the most efficient SH conversion was collected and is shown in Figure 3b. From Figure 3b, it is seen that the pump light is close to the critical coupling state, and the loaded and intrinsic quality factors of the TFLN resonator were measured to be 4.7×10$^5$ and 9.4×10$^5$, respectively. When the on-chip power of the pump is 6.3 μW, the generated SH power was measured to be 154 nW, indicating a normalized (absolute) conversion efficiency of 3.9 ×10$^5$% W$^{-1}$ (2.4%). Here, the normalized (absolute) SH conversion efficiency is defined as $\eta_{\text{norm}}=P_{\text{SH}}/P_{\text{FH}}^2 \cdot 100\%$ ($\eta=P_{\text{SH}}/P_{\text{FH}} \cdot 100\%$), with $P_{\text{FH}}$ and $P_{\text{SH}}$ indicating the pump and signal power on the chip, respectively.

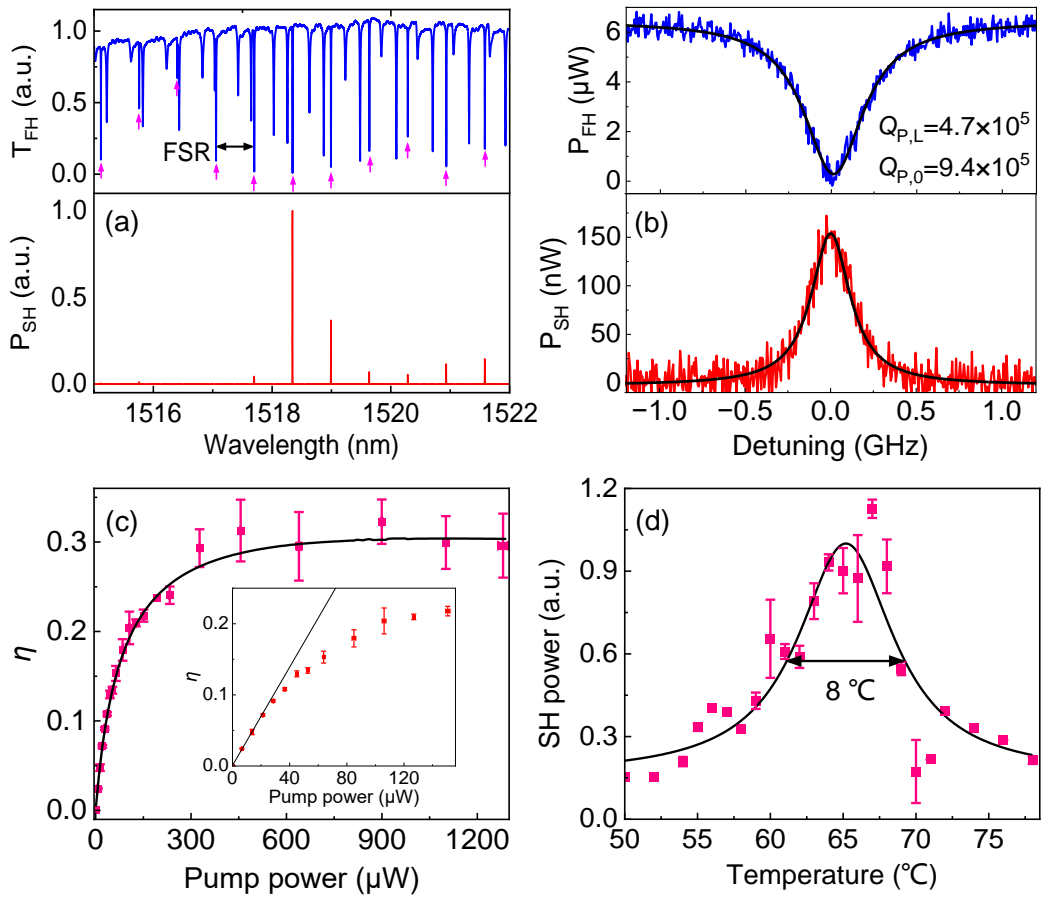

**Figure 3.** Second Harmonic Generation. a) The transmission spectrum of FH transmission and SH signal obtained by scanning the wavelength of the pump laser. b) The scanning signal of the piezo-controlled pump laser, the pump is basically in a critical coupling state, and the intrinsic Q value is above $9.4\times10^5$. The conversion efficiency was calculated as $3.9\times10^5$% $W^{-1}$ under a 6.3 μW pump. c) The SH conversion efficiency varies with the pump power. The inset shows the conversion efficiency at the low pump power. d) Temperature-tuned SH showing a bandwidth of 8 °C.

We then changed the pump power and measured the SH conversion efficiency. The corresponding results are shown in Figure 3c, where the error mainly comes from the coupling between the waveguide and the lensed fiber. By fitting the experimental data according to the coupled mode theory equations,[19] we obtained a normalized SH conversion efficiency of $4.4\times10^5$% $W^{-1}$ under very weak pumping condition. The inset of Figure 3c shows a linear fitting of the data collected with $P_{FH}<14$ μW, which is in line with the theoretical conclusion, at low pump power, the output power of the SH is quadratic with the pump and indicated $\eta_{norm}= 3.5\times10^5$% $W^{-1}$. The difference between $4.4\times10^5$% $W^{-1}$, $3.9\times10^5$% $W^{-1}$, and $3.5\times10^5$% $W^{-1}$ is mainly due to more severe pump attenuation induced by the higher absolute SH conversion efficiency with increased pump power. We can see that when the on-chip pump power reaches 600 μW, the SH efficiency is close to saturation by 30%. Table 1 compares the most efficient SH in nanophotonic resonators. Our results are the best one in resonators without PPLN domain structures.

**Table 1.** Comparison of SH conversion efficiencies in microresonators

| Type | $\eta_{norm}$ [% $W^{-1}$] | $\eta_{max}$ |
|---|---|---|
| GaAs[18] | 100 | - |
| AlN[20] | $1.7\times10^4$ | 11% @35 mW |
| LN[21] | $1.5\times10^3$ | - |
| LN[22] | $9.9\times10^3$ | - |
| PPLN[23] | $2.3\times10^5$ | 6% @35.3 μW |
| PPLN[24] | $1.4\times10^5$ | 58% @3.4 mW |
| PPLN[16] | $5.0\times10^6$ | 31% @14 μW[a] |
| LN [This work] | $4.4\times10^5$ | 30% @600 μW |

[a]Estimated from Figure 2 in Reference.[16]

We measured the SH signal at different temperatures shown in Figure 3d, where the peak corresponds to the case that the FH and SH are dual-resonant. The bandwidth of the curve for the SH generation is 8 °C. The large temperature bandwidth makes the thermal-sensitive lithium niobate microcavity work stably.

## 4. Cascaded Optical Parametric Oscillation

In a higher pump power, we even observed cascaded optical parametric oscillation (OPO) generation using optical spectrum analyzer. The principle of the cascaded OPO is shown in **Figure 4**a, where the pump ($\omega_p$) drives the SH process and produces light with an angular frequency of $\omega_{SH}$. A sufficiently high-power SH can drive the OPO process generating signals near the FH. The signal of the cascaded OPO obtained with a pump power of 1.28 mW is shown in Figure 4b, where the frequency of the OPO signal and the pump differ by 11 FSR. The transmission spectrum scanning from short wave to long wave of the FH and SH at different pump power are shown in Figure 4c, from which we can estimate that the threshold of the cascaded OPO is about 1.21 mW. We calculate the theoretical threshold of cascaded OPO of 1.04 mW with experimental parameters, which is far less than the threshold of 350 mW for four-wave mixing with Kerr effects (See Supporting Information). Therefore, it can be concluded that the OPO is due to cascaded second-order nonlinear processes. The low-threshold OPO has the potential to generate Pockels optical frequency combs for soliton generation and on-chip implementation of 1f-2f comb self-reference.[1, 25, 26]

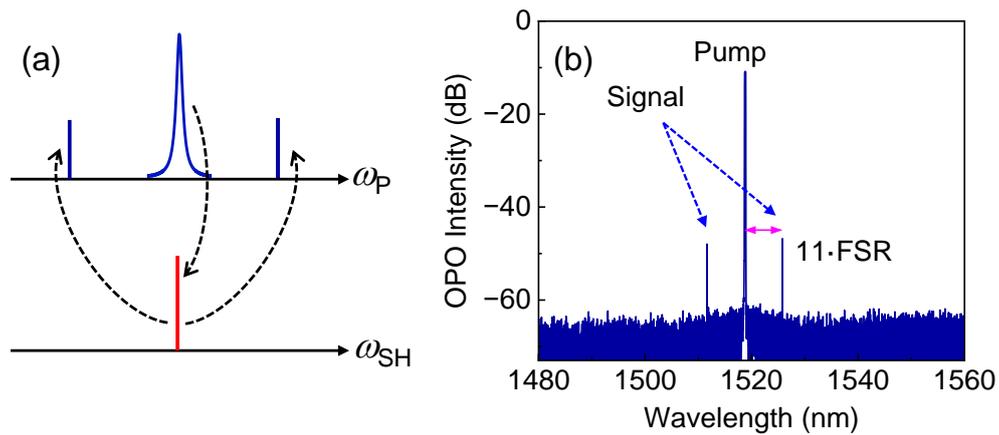

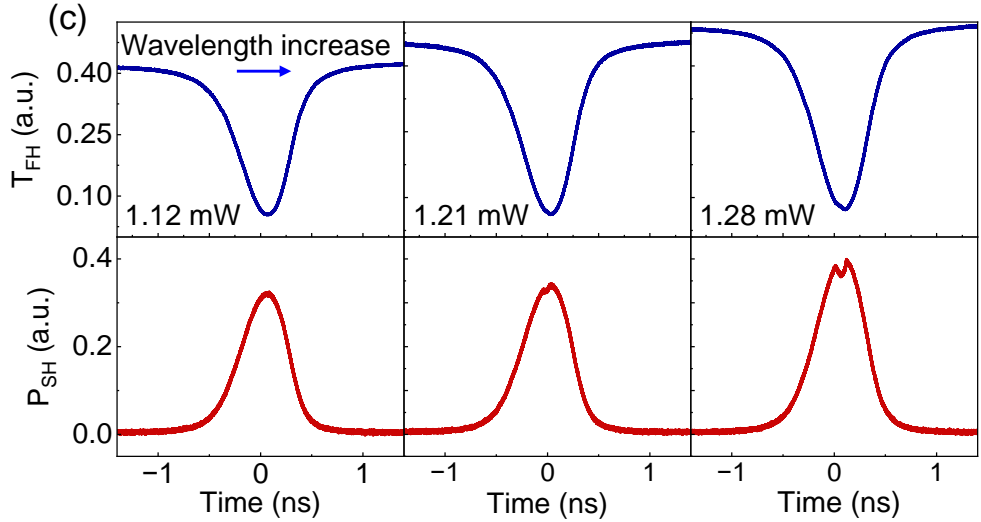

**Figure 4.** Cascaded optical parametric oscillation generation. a) Schematic diagram for cascaded OPO. b) The cascaded OPO signals at a pump of 1.28 mW. c) The transmission spectra of FH and SH at different pump powers.

## 5. Conclusion

In summary, we have generated efficient SH conversion from $TE_{00,\,FH}$ to $TE_{10,\,SH}$ in an RP dual-layer TFLN microcavity. At an optimized temperature, the normalized SH conversion efficiency reaches $4.4 \times 10^5 \%\ W^{-1}$, and the absolute conversion efficiency is up to 30% when the pump power is 600 μW. At 1.21 mW pump power, cascaded optical parametric processes were also observed due to efficient SH generation. The simple preparation process and ultra-high conversion efficiency of the RP dual-layer TFLN microresonators will facilitate various applications, such as the frequency up conversion of weak infrared signal and the generation of squeezed light and entangled photon pairs.

## Supporting Information

Supporting Information is available from the Wiley Online Library or from the author.

## Acknowledgements

X.W. and Z.H. contributed equally to this work. We would like to acknowledge Ms. Shuting Kang and Mr. Xueshan Zheng for helpful discussions when doing this work. This work was supported by the National Key Research and Development Program of China (2019YFA0705000), the National Natural Science Foundation of China (12034010, 11734009, 11674181, 11674184, 11774182, 12004197), Higher Education Discipline

Innovation Project (B07013), the National Science Fund for Talent Training in the Basic Sciences (J1103208), and PCSIRT (IRT_13R29).

**Conflict of Interest**

The authors declare no conflict of interest.

**Data Availability Statement**

The data that support the findings of this study are available from the corresponding author upon reasonable request.

Received: ((will be filled in by the editorial staff))
Revised: ((will be filled in by the editorial staff))
Published online: ((will be filled in by the editorial staff))